\documentclass[review]{elsarticle}
\usepackage{hyperref}
\usepackage{amsmath,mathtools}
\usepackage{graphicx}
\usepackage{color}
\usepackage{txfonts}
\usepackage{units}
\usepackage{amsmath}
\usepackage{amsfonts}
\usepackage{amssymb}
\newcommand{\ck}{c^{\vphantom{\dagger}}_{\! k}}
\newcommand{\ckd}{c^{\dagger}_{\! k}}

\newcommand{\dq}{d_q}
\newcommand{\e}{\mathrm e}

\newcommand{\epsk}[1]{\epsilon_{#1}}

\newcommand{\fbraket}[1]{\mathinner{\langle{#1}\rangle}}

\newcommand{\FN}{F_{\! N}}
\newcommand{\Gt}{\tilde{G}}

\newcommand{\Hop}{\hat{\mathsf H}}
\newcommand{\im}{\mathrm i}

\newcommand{\muGCE}{\bar{\muup}}

\newcommand{\Nc}{N_{\mathsf c}}

\newcommand{\Nmin}{N_{\mathsf{min}}}
\newcommand{\Nop}{\hat{\mathsf N}}
\newcommand{\ns}{n_{\mathrm S}}

\newcommand{\ZN}{Z_N}

\renewcommand{\d}{\mathrm d}
\renewcommand{\vec}[1]{\mathbf #1}
\journal{Physica A}
\begin{document}
\begin{frontmatter}
\title{Occupation numbers in a quantum canonical ensemble: a projection
       operator approach}
\author[address1,address2]{Wim Magnus\corref{correspondingauthor}}
\cortext[correspondingauthor]{Corresponding author}
\ead{wim.magnus@uantwerpen.be}
\author[address3]{Fons Brosens}
\ead{fons.brosens@uantwerpen.be}
\address[address1]{imec, Kapeldreef 75, B-3001 Leuven, Belgium}
\address[address2]{Universiteit Antwerpen, Physics Department,
                   Groenenborgerlaan 171, B-2020 Antwerpen, Belgium}
\address[address3]{Universiteit Antwerpen, Physics Department,
                   Universiteitsplein 1, B-2610 Antwerpen, Belgium}
\begin{abstract}
Recently, we have used a projection operator to fix the number of particles
in a second quantization approach in order to deal with the canonical
ensemble. Having been applied earlier to handle various problems in
nuclear physics that involve fixed particle numbers, the projector
formalism was extended to grant access as well to quantum-statistical
averages in condensed matter physics, such as particle densities and
correlation functions. In this light, the occupation numbers of the
subsequent single-particle energy eigenstates are key quantities to be
examined.
The goal of this paper is 1) to provide a sound extension of the projector
formalism directly addressing the occupation numbers as well as the
chemical potential, and 2) to demonstrate how the emerging problems related
to numerical instability for fermions can be resolved to obtain the
canonical statistical quantities for both fermions and bosons.
\end{abstract}
\end{frontmatter}

\section{Introduction}

In a previous paper~\cite{WMLLFB2017} we proposed a projection operator for
dealing with the particle number constraint in the framework of the
canonical ensemble (CE). As a result, a transparent integral representation
was obtained for the partition function $Z_N(\beta)$~
\footnote{The
interpretation of $\beta$ should be handled with care. Thermal equilibrium
means that the internal energy $U_N$ is stable in time, and $\beta$ is in
essence a Lagrange multiplier for imposing that stability, rather than a
given quantity. The internal energy 
$U_N(\beta) = -\d (\ln (Z_N(\beta))) / \d \beta$ is in fact
the fixed quantity. This correct interpretation of the principle of maximum
entropy in thermal equilibrium was treated in~Appendix A
of~\cite{WMLLFB2017}.
}
of $N$ identical indistinguishable fermions or bosons:
\begin{equation}
   Z_N(\beta
   = \frac{1}{2 \pi} \int_{-\pi}^{\pi} \! G(\beta, \theta) \, 
     \e^{-\im N \theta} \, \d \theta \quad ; \quad
   G(\beta, \theta)
   = \mathsf{Tr} \left( \e^{-\beta \Hop} \e^{\im \Nop \, \theta} \right), 
   \label{eq:Intro:Z_NinG}
\end{equation}
in which the Hamiltonian $\Hop$ and the number operator $\Nop$ are of
course assumed to be compatible operators. The trace of $G(\beta, \theta)$
is to be taken over the entire Fock space, while the angular integration
takes care of the projection onto the $N$-particle subspace.

In principle, this approach is applicable to interacting particles, but
exactly solvable systems of this kind are extremely rare.
Mostly, one has to rely on perturbational or variational treatments,
starting from non-interacting particles with supposedly known eigenstates
and energy levels.
As an example, we quote various studies~\cite{ElzeGreiner1,ElzeGreiner2,
Elzeetal,Benderetal,Landsberg,BormannFranke}
having applied the projector operator technique in a quite beneficial and
successful way.
Furthermore, the extension~\cite{WMLLFB2017} of the method enabling the
explicit calculation of correlation functions, paved the way towards
systematic explorations in condensed matter physics. A workable and
reliable algorithm yielding the particle occupation numbers (or,
equivalently, the distribution functions) is paramount in this respect,
while being the main subject of this paper.
In order to keep the course of the theory self-contained, we briefly return
to the very basics of the projector formalism.

Given a system of non-interacting bosons or fermions, the Hamiltonian
$\Hop$ and the number operator $\Nop$ can then be expressed in terms of the
single-particle energy spectrum $\epsk{k}$, where $k$ denotes any set of
generic quantum numbers properly labeling the single-particle energies:
\begin{equation}
   \Hop = \sum_k \hat{n}_k^{\vphantom{\dagger}}
          \epsk{k}^{\phantom{\dagger}} \quad ; \quad
   \Nop = \sum_k \hat{n}_{k}^{\vphantom{\dagger}} \quad ; \quad
   \hat{n}_{k}^{\vphantom{\dagger}} = \ckd \ck,
   \label{eq:Intro:H&Nin_nk}
\end{equation}
where the creation and destruction operators $\ckd$ and $\ck$ satisfy
appropriate (anti)commuta\-tion relations, i.e.
\begin{equation}
   \hat{c}_{k}^{\dagger} \hat{c}_{k^{\prime}}^{\dagger} \! -
   \xi \hat{c}_{k^{\prime}}^{\dagger}  \hat{c}_{k}^{\dagger}
   = \hat{c}_{k}^{\vphantom{\dagger}} 
     \hat{c}_{k^{\prime}}^{\vphantom{\dagger}} \! -
     \xi \hat{c}_{k^{\prime}}^{\vphantom{\dagger}}
     \hat{c}_{k}^{\vphantom{\dagger}}
   = 0 \; ; \;
     \hat{c}_{k}^{\vphantom{\dagger}}
     \hat{c}_{k^{\prime}}^{\dagger} \! -
     \xi \hat{c}_{k^{\prime}}^{\dagger}
     \hat{c}_{k}^{\vphantom{\dagger}}
   = \delta_{k,k^{\prime}} \; ; \;
   \xi =
   \begin{cases}
      +1 \!\!\! & \text{for bosons,} \\
      -1 \!\!\! & \text{for fermions.}
   \end{cases}
   \label{eq:Intro:commutators}
\end{equation}
As detailed in~\cite{WMLLFB2017}, the projector formalism enables an
unrestricted summation over the occupation numbers $n_k$ entering the
expression for $G(\beta, \theta)$:
\begin{equation}
   G(\beta, \theta)
   = \mathsf{Tr} \left( \e^{-\beta \Hop} \e^{\im \Nop \, \theta} \right)
   = \prod_k
     \left(
           \sum_{n_k}
           \exp \left( (\im \theta -\beta \epsk{k}) n_k \right)
     \right)
\end{equation}
Summing $n_{k}$ from $0$ to $\infty$ for bosons, and from $0$ to $1$ for
fermions, readily gives
\begin{equation}
   G(\beta, \theta)
   = \prod_k
     \left(
           1 - \xi \exp \left( \im \theta -\beta \epsilon_{k} \right) 
     \right)^{-\xi}.
   \label{eq:Gxi}
\end{equation}
It should be noted, however, that the geometric series~
\footnote{
Remarkably, the common ratio of a similar geometric series 
appearing in the grand-canonical partition function crucially depends on
the grand-canonical chemical potential $\muGCE(N)$. More specifically, as
convergence requires the common ratio to be smaller than 1, $\muGCE(N)$ is
bound to be located below $\epsk{0}$.
The latter, in turn, requires that the single-particle ground-state energy
be strictly positive. In this light, it is explicitly assumed that
$\epsk{0} > 0$ until the recurrence relations for the partition function
and the occupation numbers are established. Afterwards, a simple gauge
transformation consisting of a constant energy shift can be performed to
generalize the results to the case of arbitrary, but finite values of
$\epsk{0}$.
}
leading to
(\ref{eq:Gxi}) for bosons ($\xi = +1$), only converges if
$|\exp (\im \theta -\beta \epsk{k})| < 1$ holds for all $k$.
The angular integration can equivalently be expressed as a complex contour
integral along a circle with radius $r$ enclosing the origin:
\begin{equation}
   Z_{N}(\beta)
   = \frac{1}{2 \pi \im}
     \oint_{\left \vert z \right \vert = r}
     \frac{\tilde{G} \left( \beta, z \right)}{z^{N+1}} \, \d z \quad ;
     \quad
   \tilde{G}(\beta, z)
   = \prod_{k} \left( 1 - \xi z \e^{-\beta \epsilon_{k}}\right)^{-\xi}.
   \label{eq:Intro:Z_NinGtilde}
\end{equation}
The radius $r$ should be chosen small enough to ensure that the contour
$\left \vert z \right \vert = r$ does not enclose any of the poles of
$\tilde{G}(\beta, z)$ appearing in the case of bosons.
Though being a useful starting point for further investigations, the
above integral representations do not generally lead to
closed form expressions for $\ZN$ or quantities derived from it. As an 
exception, we mention the special case of one-dimensional harmonic
oscillators~
\footnote{
Formula (25) in~\cite{WMLLFB2017} contains a serious misprint, and should
read
\begin{equation}
   \ZN(\beta) =
   \frac{1}{\prod_{k = 1}^{N}
   \left( 1 - \e^{-k \beta \hbar \omega} \right)} \times
   \begin{cases}
      \e^{-N   \beta \hbar \omega / 2} & ~\text{for bosons,} \\
      \e^{-N^2 \beta \hbar \omega / 2} & ~\text{for fermions.}
   \end{cases}
   \tag{25}
\end{equation}
%
}
that was solved analytically upon invoking two Euler
identities~\cite{WMLLFB2017}. Unfortunately, we overlooked the magisterial
treatment of non-interacting fermions with equidistant single-particle
energies by Sch\"onhammer~\cite{Schoenhammer2000}, that turns out to
remain quite relevant to the present paper.

Although the projection operator approach was applied to derive generic
expressions for the two- and four-point correlation functions, no detailed
explicit results were reported in~\cite{WMLLFB2017}.
In section (\ref{sec:Occup}) we derive numerically tractable recurrence
relations for both the chemical potentials and the occupation numbers, the
latter being needed crucially to compute the correlation functions. In the
same section we remedy the numerical instabilities that were prohibitive
for extending the number of particles at will in the case of
fermions~\cite{WMLLFB2017}. In particular, new results are presented
addressing not only the occupation numbers but also the dependence of the
chemical potential, the Helmholtz free energy, the internal energy and the
entropy of the two-dimensional electron gas (2DEG) on the particle number.
 
\section{Occupation numbers and chemical potential \label{sec:Occup}}
Consider the occupation number $g_{k, N}(\beta)$, defined as the
expectation value 
$\left \langle \hat{c}_{k}^{\dagger} \hat{c}_{k}^{\vphantom{\dagger}}
\right \rangle_{\beta, N}$ of the $N$-particle
system~(\ref{eq:Intro:H&Nin_nk}):
\begin{equation}
   g_{k, N}(\beta)
   = -\frac{1}{\beta} \frac{1}{\ZN(\beta)} 
     \frac{\partial \ZN(\beta)}{\partial \epsk{k}}.
\end{equation}
Temporarily disregarding the trivial result $g_{k, N = 0}(\beta) = 0$, one
readily obtains from the representation~(\ref{eq:Intro:Z_NinGtilde})
\begin{equation}
   g_{k, N}(\beta)
   = \frac{\e^{-\beta \epsk{k}}}{\ZN(\beta)} \frac{1}{2 \pi \im}
     \oint_{\left \vert z \right \vert = r > 0}
     \frac{\tilde{G}(\beta, z)}{1 - z \xi \e^{-\beta \epsk{k}}}
     \frac{1}{z^N} \, \d z.
     \label{eq:Occup:g_in_Gtilde}
\end{equation}
Because of the pole of order $N$ in the origin, the residue theorem yields
\begin{equation}
   g_{k, N}(\beta)
   = \frac{1}{\ZN(\beta)} \frac{\e^{-\beta \epsk{k}}}{(N - 1)!}
   \left. 
         \frac{\partial^{N - 1}}{\partial z^{N - 1}}
         \frac{\tilde{G}(\beta, z)}{1 - \xi z \e^{-\beta \epsk{k}}}
   \right\vert_{z = 0}.
\end{equation}
Using 
$\frac{\partial^j}{\partial z^j} \frac{1}{1-az} =
\frac{j! a^j}{(1 - a z)^{j + 1}}$ and 
$\left. \frac{\partial^n \tilde{G}(\beta, z)}{\partial z^n}
\right \vert_{z = 0} = n! \, Z_{n}(\beta)$ 
in Leibniz' differentiation rule for function products, one ends up with
\begin{equation}
   g_{k, N}(\beta)
   = \sum_{j = 1}^N \xi^{j - 1} \e^{-j \beta \epsk{k}}
     \frac{Z_{N - j}(\beta)}{\ZN(\beta)}.
\end{equation}
Separating the first term ($j = 1$) and substitution $j \to j - 1$ into
the remaining sum, one immediately recognizes a recurrence relation
\begin{equation}
   g_{k, N}(\beta)
   = \left( 1 + \xi g_{k, N - 1}(\beta) \right) \e^{-\beta \epsk{k}}
     \frac{Z_{N - 1}(\beta)}{\ZN(\beta)},
\end{equation}
that was earlier obtained by Schmidt~\cite{Schmidt1953} and exploited by
Sch\"onhammer to treat fermionic systems $(\xi = -1)$ (see Eq.~(19)
of~\cite{Schoenhammer2000}).

If $\epsk{k = 0}$ had to be shifted to a positive value in order to avoid
spurious poles in the complex plane, one might choose to undo the
corresponding gauge transformation at this point since all complex
integrations required to set up the recurrence relation are carried out.

Introducing the standard definition of the chemical
potential~\footnote{In Eq.~(12) of~\cite{Schoenhammer2000}
$F_N - F_{N - 1}$ was used to define $\muup_N$, rather
than Eq.~(\ref{eq:Occup:mu_def}) in the current paper} in the CE,
\begin{equation}
   \muup_N(\beta) = F_{N + 1}(\beta) - F_N(\beta) \quad
   \text{with} \;\, \ZN(\beta) = \e^{-\beta F_N(\beta)},
   \label{eq:Occup:mu_def}
\end{equation}
and using $\sum_k g_{k, N}(\beta) = N$, one obtains
\begin{align}
   g_{k, N}
   & = \e^{-\beta(\epsk{k} - \muup_{N - 1})}
       \left( 1 + \xi g_{k, N - 1} \right),
       \label{eq:Occup:g_recur} \\
   \e^{-\beta \muup_{N-1}}
   & = \frac{1}{N} \sum_k \e^{-\beta \epsk{k}}
       \left( 1 + \xi g_{k, N - 1} \right),
       \label{eq:Occup:mu_recur}
\end{align}
where the temperature parameter $\beta$ (considered to be fixed for the
time being) was omitted as an argument for the sake of notation's
simplicity in the subsequent calculations. The initialization of the above
recursion is simple:
\begin{equation}
   g_{k, 0} = 0 \quad ; \quad
   \e^{-\beta \muup_0} = \sum_k \e^{-\beta \epsk{k}}.
   \label{eq:muzero_def}
\end{equation}
Note that $\e^{\beta \muup_{N - 1}}$ is the basic numerical quantity for
implementing the recurrence. At the end of the calculations the chemical
potential itself and, hence, also the free energy can be easily obtained.

Anticipating the numerical implementation, the relation to the grand
canonical ensemble (GCE) may be beneficially established at this
point by comparison with the distribution function of the GCE, i.e.
\begin{equation}
   f_k(\muup) = \frac{1}{\e^{\beta(\epsk{k} - \muGCE_N)} -\xi}
                \quad ; \quad \sum_k f_k(\muGCE_N) = N,
                \label{eq:Occup:f_def}
\end{equation}
where $\bar{\muup}_N$ (to be distinguished from $\muup_N)$ denotes the
chemical potential in the GCE.

For arbitrary values of $N$ and $N^{\prime}$ the relation between
$g_{k, N - 1}$ and $f_k(\bar{\muup}_{N^{\prime}})$ can be further
elucidated.  Assuming the validity of the generic inequality,
$0\leq g_{k, N - 1} \leq g_{k, N}$, the recursion~(\ref{eq:Occup:g_recur})
immediately implies
\begin{equation}
   g_{k, N - 1} \leq f_k(\muup_{N - 1}).
   \label{eq:Occup:g_N<f_N}
\end{equation}
Furthermore, the identity 
$1 = (\e^{\beta (\epsk{k} - \muup_{N - 1})} - \xi) f_k(\muup_{N - 1})$
allows to replace $1$ in the factor $(1 + \xi g_{k, N - 1})$
of~(\ref{eq:Occup:g_recur}). This leads to an alternative formulation of
the original recurrence relation:
\begin{equation}
   g_{k, N}
   = f_k(\muup_{N - 1}) + \xi \e^{-\beta (\epsk{k} - \muup_{N - 1})}
     \left( g_{k, N - 1} - f_k(\muup_{N - 1}) \right). 
   \label{eq:Occup:g&f_recur}
\end{equation}
Combining (\ref{eq:Occup:g&f_recur}) with~(\ref{eq:Occup:g_N<f_N}) for
fermions, we may infer $g_{k, N} \geq f_k(\muup_{N - 1})$, thus arriving
at
\begin{equation}
   f_k(\muup_0) \leqslant g_{k, 1} \leqslant \cdots \leqslant
   f_k(\muup_{N - 1}) \leqslant g_{k, N} \leqslant f_k(\muup_N)
   \leqslant \cdots \quad \text{for fermions.}
   \label{eq:Occup:g<fFermi}
\end{equation}
For bosons $(\xi = +1)$ we were unable to find a similar ladder relation,
but the inequality~(\ref{eq:Occup:g_N<f_N}) can now be replaced by a
stronger one:
\begin{equation}
   g_{k, N} \leqslant f_k(\muup_{N - 1})
   \quad \text{for bosons.}
   \label{eq:Occup:g<fBose}
\end{equation}
At this point, the simultaneous treatment of bosons and fermions becomes a
hindrance rather than a convenience and, hence, we treat fermions and
bosons separately from hereof.

\section{Boson occupation numbers\label{sec:Bose}}

For bosons ($\xi = +1$), the recurrence relations~(\ref{eq:Occup:g_recur})
and~(\ref{eq:Occup:mu_recur}) now become
\begin{align}
   g_{k,N}  
   & = \e^{-\beta (\epsk{k} - \muup_{N - 1})}
       \left( 1 + g_{k, N - 1} \right),
       \label{eq:Bose:g_recur} \\
   \e^{-\beta \muup_{N - 1}} 
   & = \frac{1}{N} \sum_k \e^{-\beta \epsk{k}}
       \left( 1 + g_{k, N - 1} \right).
       \label{eq:Bose:mu_recur}
\end{align}
As commonly known, the GCE chemical potential that fixes the
{\itshape average} number of particles rather than the actual, integer
number of particles, does not exceed the single-particle ground state
energy $\epsilon_{0}$. Accordingly, it is quite tempting to consider
$\epsilon_{0}$ as well as a rigorous upper bound for any $\muup_N$,
although the formal proof turns out to be less trivial than in the GCE
case (see \ref{app:muNvseps0}).
The restriction $\muup_N < \epsilon_0$ ensures the numerical stability of
the encoded recurrence relations
(\ref{eq:Bose:g_recur}--\ref{eq:Bose:mu_recur}), although one may have
to remedy some overflow and underflow deficiencies appearing in the case of
extremely low temperatures.

As an example, we treat bosonic harmonic oscillators, omitting however the
vacuum energy for the sake of simplicity in the subsequent numerical work.
If desired, it can be restored at the end of the calculations.
Accordingly, we consider the Hamiltonian
\begin{equation}
   \Hop = \sum_{\vec{k}} \epsilon_{\vec{k}}^{\phantom{\dagger}}
          c_{\vec{k}}^{\dagger} c_{\vec{k}}^{\phantom{\dagger}}
\end{equation}
where the components of $\vec{k}$ are non-negative integers and the 
energy spectrum is given by
\begin{equation}
   \epsilon_{\vec{k}}^{\phantom{\dagger}}
   = \hbar \omega \sum_{j = 1}^{D} k_j \quad ; \quad k_j = 0, 1, \cdots,
     \infty.
\end{equation}
$D$ is the spatial dimension (1, 2 or 3 at will).
In view of the rapidly growing degeneracy, its proves more natural to
relabel the single-particle energy levels in terms of a shell index $q$,
pointing to an energy shell of states that share a common energy
$\tilde{\epsilon}_{q}$ with a dimension dependent degeneracy $\dq$:
\begin{equation}
   q = 0, 1, 2, 3, \ldots \quad ; \quad
   \tilde{\epsilon}_{q} = \hbar \omega q \quad ; \quad
   \dq = 
   \begin{cases}
                       1 & \text{if $D = 1$}, \\
                   q + 1 & \text{if $D = 2$}, \\
      (q + 1)(q + 2) / 2 & \text{if $D = 3$}.
   \end{cases}
\end{equation}
The occupation numbers $g_{\vec{k},N}$ having the same degeneracy as
$\epsilon_{\vec{k}}$, $\tilde{g}_{q,N}$ denotes the occupation number of
any particular state in the energy shell $\tilde{\epsilon}_{q}$,
that, in turn, determines the occupation probability $p_{q,N}$ of any
energy level in the shell $\tilde{\epsilon}_{q}$: 
\begin{equation}
   p_{q, N} = \frac{1}{N} \dq \tilde{g}_{q,N}.
\end{equation}
The recurrence relations~(\ref{eq:Bose:g_recur}--\ref{eq:Bose:mu_recur})
for $\tilde{g}_{q, N}$ become
\begin{align}
   \tilde{g}_{q,N}  
   & = \e^{-\beta (\hbar \omega q - \muup_{N - 1})}
       \left( 1 + \tilde{g}_{q, N - 1} \right), \\
   \e^{-\beta \muup_{N - 1}}  
   & = \frac{1}{N} \sum_{q = 0}^{\infty} \dq \e^{-\beta \hbar \omega q} 
       \left( 1 + \tilde{g}_{q, N - 1} \right),
\end{align}
and are now initialized by
\begin{equation}
   \tilde{g}_{q, N = 0}
   = 0 \quad; \quad
   \tilde{g}_{q, N = 1}
   = \e^{-\beta \hbar \omega q} \e^{\beta \muup_{0}}
   = \e^{-\beta \hbar \omega q}
     \left( 1 - \e^{-\beta \hbar \omega} \right)^D,
\end{equation}
where the required summation in
$\muup_0(\beta) = -(1 / \beta) \sum_k \e^{-\beta \epsk{k}}$, as appearing
in Eq.~(\ref{eq:muzero_def}), was done analytically.

In one dimension ($D = 1$), one can check by induction that
\begin{align}
   \left. \tilde{g}_{q, N} \right\vert_{\mathrm{1D}}
   & = \sum_{j = 1}^N \e^{-\beta \hbar \omega q j}
       \prod_{m = N + 1 - j}^N
       \left( 1 - \e^{-\beta \hbar \omega m} \right), \\
   \left. \e^{\beta \muup_N} \right\vert_{\mathrm{1D}}
   & = 1 - \e^{-\beta \hbar \omega (N + 1)},
\end{align}
which is useful to monitor the numerical recursion work, because it is
tractable with symbolic algebra with only 2 independent parameters, namely
$N$ and the dimensionless temperature $\tau = 1 / (\beta \hbar \omega)$.

\begin{figure}[ht]
   \centering{
   \includegraphics[width = 0.85\textwidth]{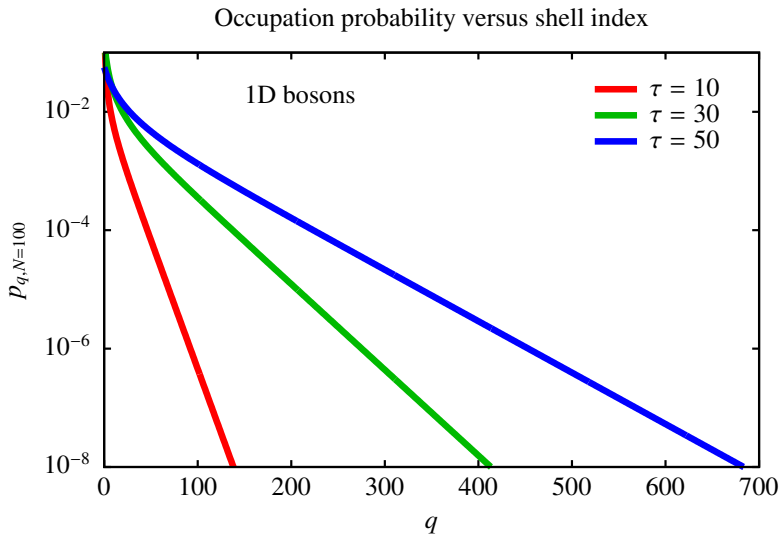}
   \caption{Probability $p_{q, N}$ of occupying energy level
            $\tilde{\epsilon}_{q}$ for 100 1D harmonic oscillators as
            a function of the shell index $q$, given three values of the
            dimensionless temperature $\tau = 1 / (\beta \hbar \omega)$.
           }
   \label{Fig:p_N100_D1_tauFX_q0-qmax}
             }
\end{figure}

\begin{figure}[ht]
   \centering{
   \includegraphics[width = 0.85\textwidth]{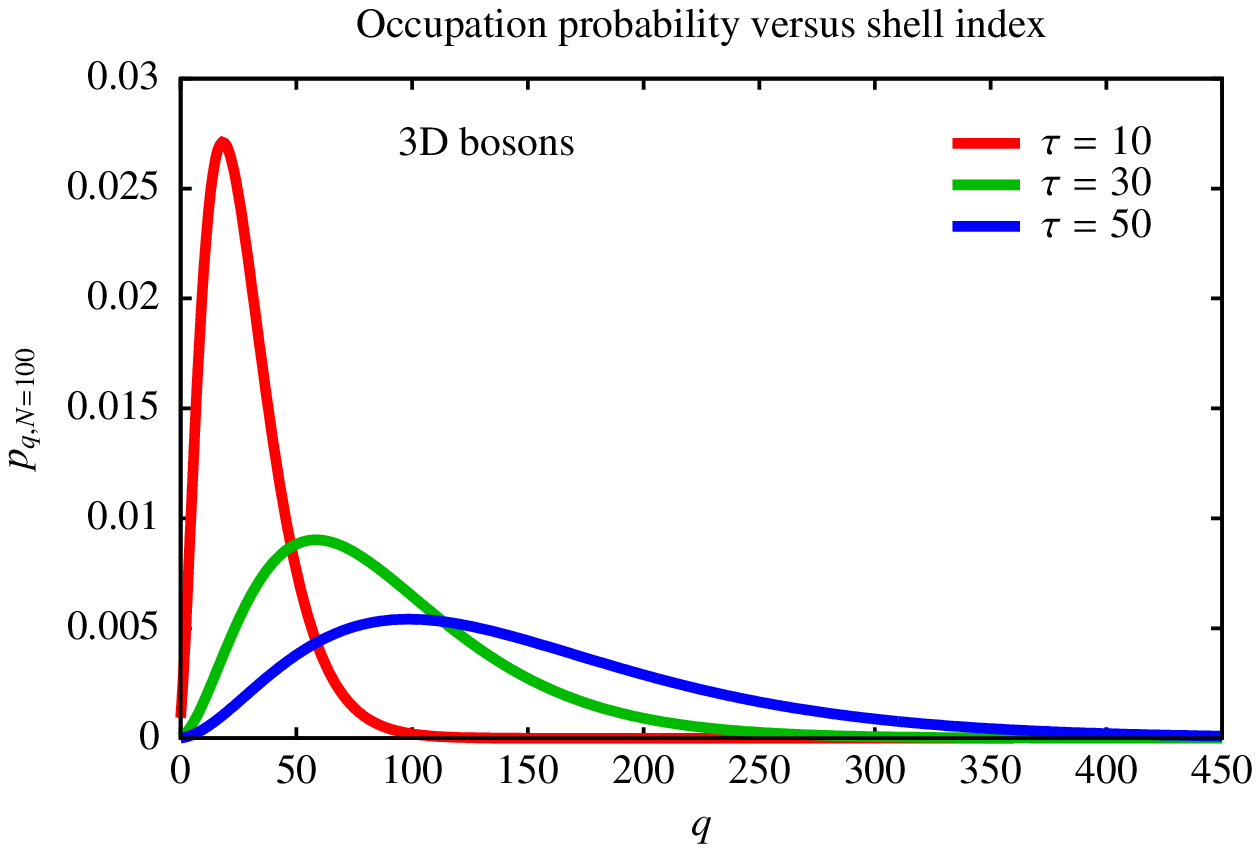}
   \caption{Probability $p_{q, N}$ of occupying energy level
            $\tilde{\epsilon}_{q}$ for 100 3D harmonic oscillators as
            a function of the shell index $q$, given three values of the
            dimensionless temperature $\tau = 1 / (\beta \hbar \omega)$.
           }
   \label{Fig:p_N100_D3_tauFX_q0-qmax}
             }
\end{figure}

\begin{figure}[ht]
   \includegraphics[width = 0.85\textwidth]{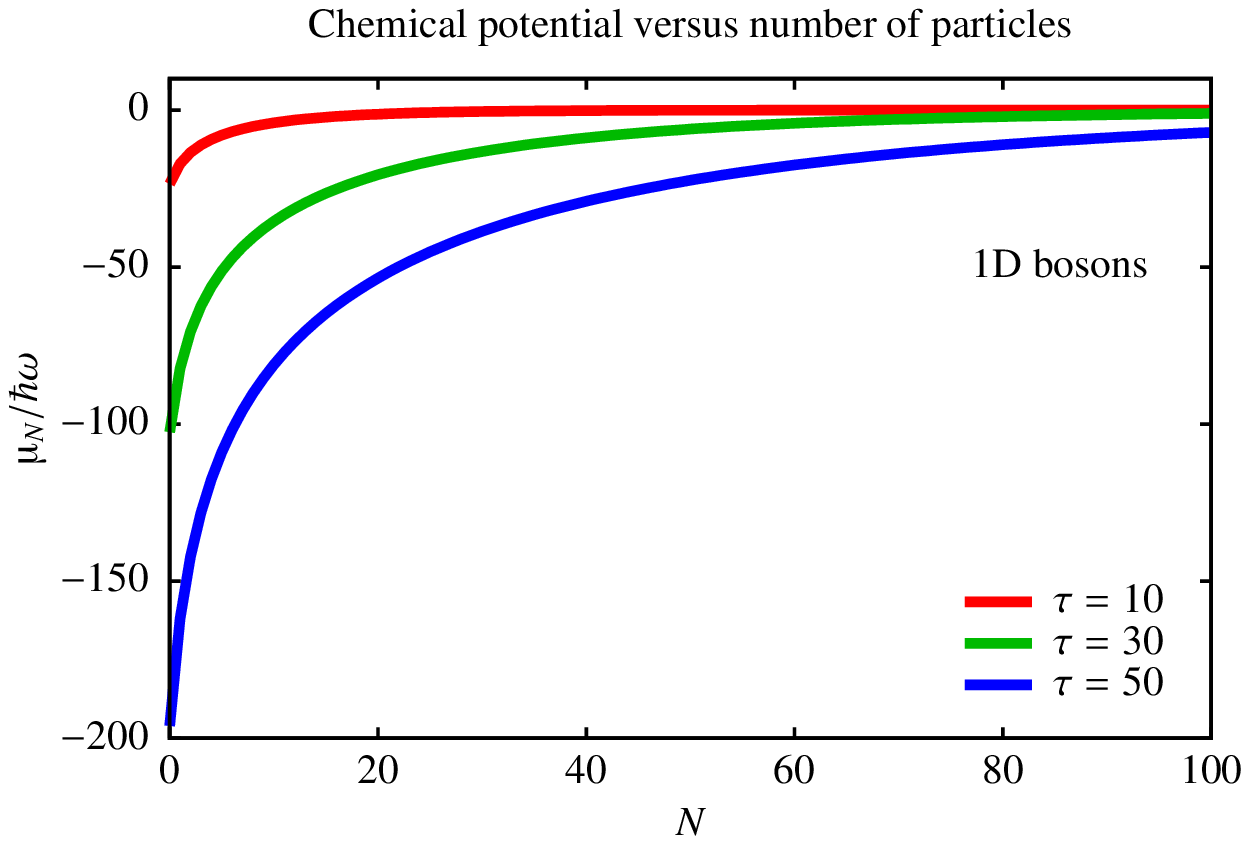}
   \centering{
   \caption{Scaled chemical potential $\muup_N / (\hbar \omega)$ for 1D
            bosonic harmonic oscillators as a function of the number of
            particles, given three values of the dimensionless temperature
            $\tau = 1 / (\beta \hbar \omega)$.
            }
   \label{Fig:mu_tauFX_D1_N0-100}
             }
\end{figure}

\begin{figure}
   \centering
   \includegraphics[width = 0.85\textwidth]{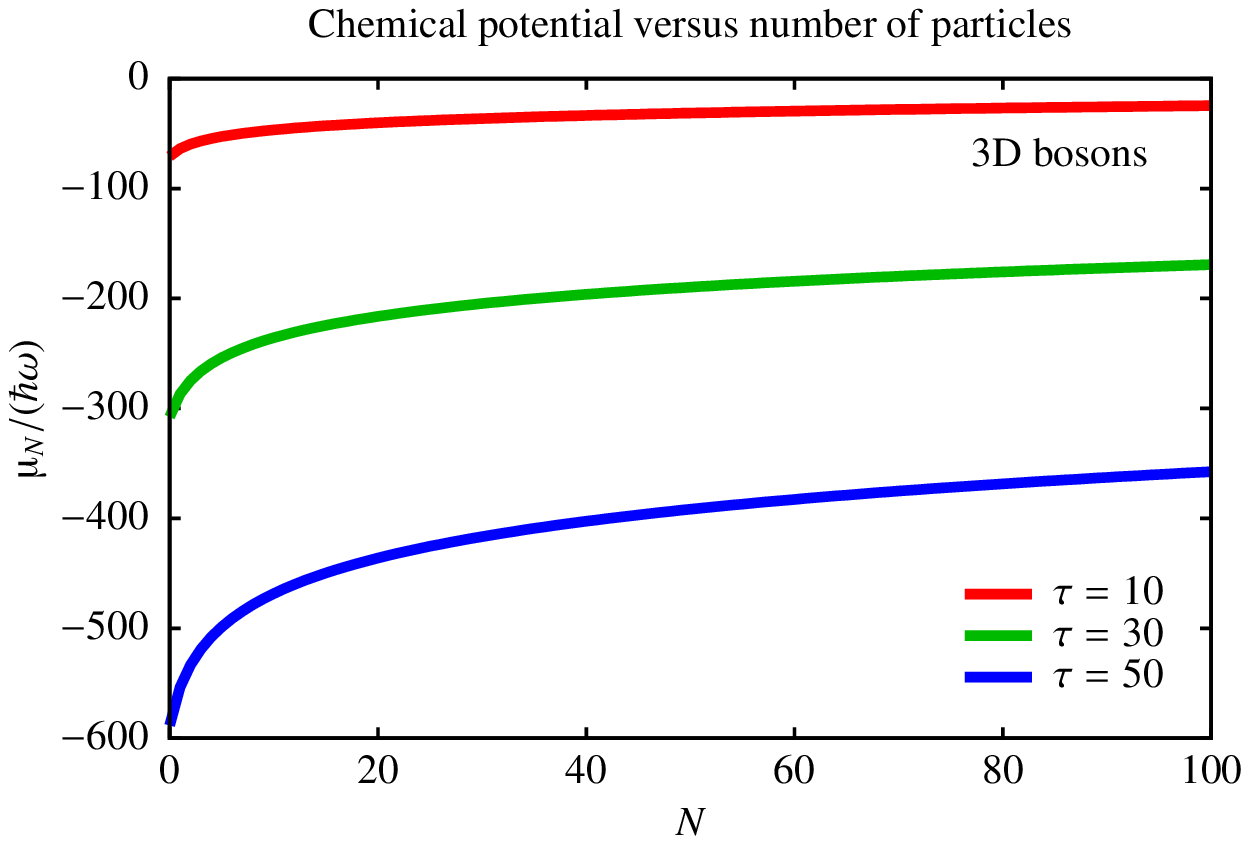}
   \caption{Scaled chemical potential $\muup_N / (\hbar \omega)$ for 3D
            bosonic harmonic oscillators as a function of the number of
            particles, given three values of the normalized temperature
            $\tau = 1 / (\beta \hbar \omega)$.}
   \label{Fig:mu_tauFX_D3_N0-100}
\end{figure}

For 100 oscillators the energy occupation numbers for the 1D case and the
3D case are shown in Fig.~(\ref{Fig:p_N100_D1_tauFX_q0-qmax}) and
Fig.~(\ref{Fig:p_N100_D3_tauFX_q0-qmax}).
Apparently, the profiles of the latter case attain maximum levels as a
direct consequence of the non-trivial 3D degeneracy factors $\dq$.
The evolution of the chemical potential (in units of $\hbar \omega)$ as a
function of the number of particles is shown in
Fig.~(\ref{Fig:mu_tauFX_D1_N0-100}) for 1D, and in
Fig.~(\ref{Fig:mu_tauFX_D3_N0-100}) for 3D.

\begin{figure}
   \centering
   \includegraphics[width = 0.85\textwidth]{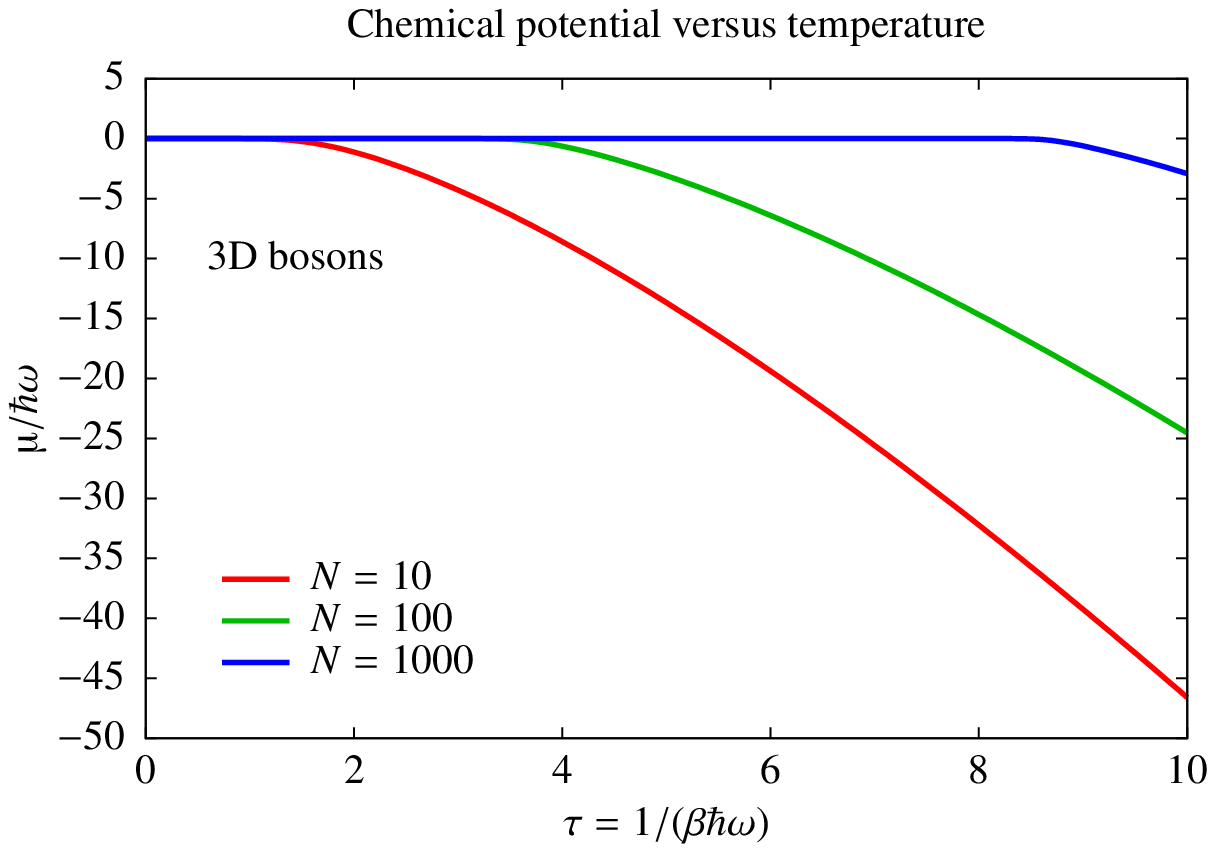}
   \caption{Scaled chemical potential $\muup_N / (\hbar \omega)$ for 100 3D
            bosonic harmonic oscillators as a function of the number of the
            normalized temperature $\tau = 1 / (\beta \hbar \omega)$.}
   \label{fig:mu_NFX_D3_tau0-10}
\end{figure}

\begin{figure}
   \centering
   \includegraphics[width = 0.85\textwidth]{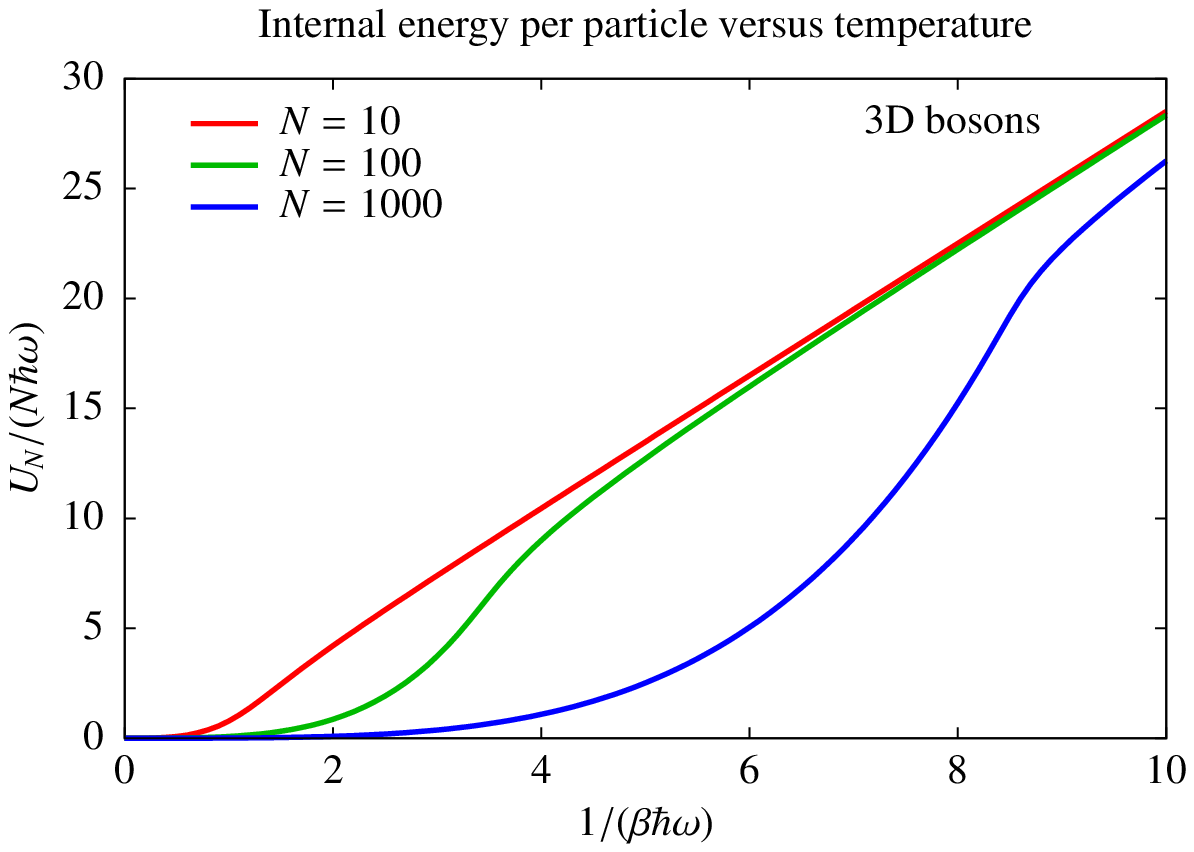}
   \caption{Scaled internal energy $U / (N \hbar \omega)$ for 3D bosonic
            harmonic oscillators as a function of the normalized
            temperature $\tau = 1 / (\beta \hbar \omega)$.}
   \label{fig:U_NFX_D3_tau0-10}
\end{figure}

Furthermore, as an illustration, we have shown the temperature dependence
of the normalized chemical potential and the internal energy
$U_N = \fbraket{\Hop}_N$ for 10, 100 and 1000 3D harmonic oscillator bosons
in Fig.~(\ref{fig:mu_NFX_D3_tau0-10}) and Fig.~(\ref{fig:U_NFX_D3_tau0-10})
respectively. As expected, the bosonic nature is most pronounced in the
low temperature regime ($\tau < 4$), whereas the internal energy matches
the classical limit, i.e.
$U_N(\tau) \to 3 N \hbar \omega / (\exp(-1 / \tau) - 1)$ at high
temperatures.
\section{Fermion occupation numbers\label{sec:Fermi}}

For fermions $(\xi = -1)$ the recurrence
relations~(\ref{eq:Occup:g_recur}--\ref{eq:Occup:mu_recur}) obviously read
\begin{align}
   g_{k, N}
   & = \e^{-\beta \left( \epsk{k} -\muup_{N - 1} \right)}
       \left( 1 - g_{k, N - 1} \right), \label{eq:Fermi:g_recur} \\
   \e^{-\beta \mu_{N - 1}}
   & = \frac{1}{N} \sum_k
       \e^{-\beta \epsk{k}} \left( 1 - g_{k, N - 1} \right).
       \label{eq:Fermi:mu_recur}
\end{align}
For the sake of convenience but without loss of generality, we may assume
that $k$ exclusively runs through non-negative integers labeling the energy
eigenvalues $\epsk{k}$ in ascending order and starting at $\epsk{0} = 0$.

The recursive solution of (\ref{eq:Fermi:g_recur}) and
(\ref{eq:Fermi:mu_recur}) is prone to numerical errors that propagate with
$k$, while being proportional to
$\e^{-\beta (\epsk{k} - \muup_{N - 1})}$. Dealing with fermions, however,
we must abandon the requirement that the chemical potential be restricted
to values below $\epsk{0} = 0$, and realize that the sign of
$\epsk{k} - \muup_{N - 1}$ determines the magnitude of the numerical errors
appearing in the recursive flow. Clearly, the error level decreases
provided that $\muup_{N - 1} < \epsk{k}$ for all $k$, i.e. as long as
$\muup_{N - 1} < 0$. For sufficiently large $N$, however, the Helmholtz
free energy attains a minimum, say at $N = \Nmin$, beyond which 
$\FN > F_{\! N - 1}$ and, hence, $\muup_{N - 1} > 0$ holds. Phrased otherwise,
once $N > \Nmin$, the chemical potential crosses the energy spectrum and,
in particular, the low energy section below $\muup_{N - 1}$ causes the
errors to grow exponentially.
Moreover, the expression for $\e^{-\beta \muup_{N - 1}}$
in~(\ref{eq:Fermi:mu_recur}) shows that $\muup_{N - 1}$ greedily
accumulates the numerical errors on $g_{k, N - 1}$. This explains in depth
the numerical accuracy problem encountered in~\cite{WMLLFB2017} where the
recurrence relation for the partition function was directly addressed.

For a typical 2DEG at room temperature and contained in a
rectangle~$0 \leqslant x \leqslant L_x$, $0 \leqslant y \leqslant L_y$, we
found that $\Nmin = 374$ for $L_x = L_y = 100$ nm. Bearing the latter in
mind as well as the gradual deterioration of the results for $g_{k, N}$ and
its cumulative effect on $\muup_{N - 1}$, to be expected when $N$ exceeds
$\Nmin$, one may now understand why the results become totally unreliable
and numerically unstable for $N > 520$ (even yielding negative values for
the partition function). Being attributed loosely to the infamous sign
problem for fermions~\cite{WMLLFB2017}, this issue is now clarified in
greater detail by the error analysis of
(\ref{eq:Fermi:g_recur}--\ref{eq:Fermi:mu_recur}).

Fortunately, the narrow boundaries~(\ref{eq:Occup:g<fFermi}) allow to
detect and correct the misbehaviour of $\muup_{N - 1}$ at an early stage if
the temperature is not extremely low. (In that case a suitable Sommerfeld
expansion might be appropriate.) When the inequality
$g_{k, N - 1} < f_k(\muup_{N - 1})$ is violated for the first time at
$k = 0$, the relative error on $\e^{-\beta \muup_{N - 1}}$ is still small.
Since this happens for $N \gg 1$, we may anticipate the detrimental
accumulation of numerical errors by exploiting the observation that the
canonical distribution function converges to the grand-canonical one, when
$N$ grows arbitrarily large. Full knowledge of $f(\muGCE)$, however, would
require us to solve the transcendental equation
$\sum_k f_k(\muGCE(N)) = N$ for $\muGCE = \muGCE(N)$, the grand-canonical
chemical potential compatible with $N$ particles. However, a first order
Taylor expansion
\begin{align}
   f_k(\muup)   
   & =   f_k(\muGCE) + \left( \muup - \muGCE \right)
         \left. \frac{\d f_k(\muup)}{\d \muup} \right|_{\muup = \muGCE}
         \!\! + {\cal O} \left( \left( \muup - \muGCE \right)^2 \right)
         \nonumber \\
   & \to f_k(\muup) \approx f_k(\muGCE) + \beta
         \left( \muup - \muGCE \right)
         \e^{\beta \left( \epsk{k} - \muGCE \right)}
         \left( f_k(\muGCE) \right)^2
   \label{eq:Fermi:f_O(1)}
\end{align}
provides a sufficiently accurate approximation in most circumstances.

Let $\Nc$ be the lowest value of $N$ for which the preceding occupation
number $g_{k = 0, \Nc - 1}$ erroneously exceeds
$f_{k = 0}(\muGCE(\Nc - 1))$ in the course of the recursion.
Imposing the sum rule $\sum_k f_k(\muup_{\Nc - 1}) = \Nc - 1$
in~(\ref{eq:Fermi:f_O(1)}) then gives the correction
\begin{equation}
   \beta \muup_{\Nc - 1}
   = \beta \muGCE(\Nc - 1) +
     \frac{\Nc - 1 - \sum_k f_k(\muGCE(\Nc - 1))}
          {\sum_k \e^{\beta (\epsk{k} - \muGCE(\Nc - 1))}
           \left( f_k(\muGCE(\Nc - 1)) \right)^{2}}.
\end{equation}
For $N > \Nc$, the recursion~(\ref{eq:Fermi:g_recur}) of course becomes
increasingly inaccurate, but since this is a regime of slowly varying
$\muup_N$ with $N$, we keep using~(\ref{eq:Fermi:f_O(1)}), with
$\tilde{\mu}$ and $\mu$ being replaced respectively by $\mu_{N - 1}$ and
$\muup_N$. Summing over $k$ one thus finds
\begin{equation}
   N > \Nc~:~\beta \muup_N = \beta \muup_{N - 1} + 
   \frac{1}{\sum_k \e^{\beta (\epsk{k} - \muup_{N - 1})}
            \left( f_k(\muup_{N - 1}) \right)^{2}}.
   \label{eq:pseudoGCE}
\end{equation}
At any stage of the calculation one easily monitors the quality of the
approach by checking whether $\sum_k g_k(\muup_N) = N$ remains valid.
In case of failure however, we have no alternative approach available so
far, and we are left with the fermion sign problem remaining prohibitive
for that particular case.

In order to test the procedure, we first apply it to a two-dimensional (2D)
electron gas for which Sch\"onhammer~\cite{Schoenhammer2000} has
developed an alternative approach by linearizing its energy spectrum.
Comparison with our approach (see below) shows an excellent agreement.

\section{Two-dimensional electron gas\label{sec:2Delgas} -- linearized
         energy spectrum}

Consider again a 2D electron gas in a rectangle
$0 \leqslant x \leqslant L_x$, $0 \leqslant y \leqslant L_y$, with periodic
boundary conditions imposed on the single-electron wave functions.
Before linearization, the energy spectrum~(\ref{eq:Intro:H&Nin_nk})
is expressed in terms of 2D  wave vectors
$k_{x, y} = 2 \pi n_{x, y} / L_{x, y}$ as
\begin{equation}
   \epsk{k} \to \epsilon_{n_x, n_y}
   = \frac{\hbar^2}{2 m_{e}}
     \left[
           \left( \frac{2 \pi n_x}{L_x} \right)^2 +
           \left( \frac{2 \pi n_y}{L_y} \right)^2
     \right]
     \quad ; \quad
     n_x, n_y = 0, \pm 1, \pm 2, \ldots \label{eq:2DEGqspectrum}
\end{equation}
where $m_{e}$ denotes the electron effective mass. Since we are dealing
with fermions $(\xi = -1)$, the recurrence
relations~(\ref{eq:Occup:g_recur}--\ref{eq:Occup:mu_recur}) obviously read
\begin{equation}
   g_{k, N}
   = \e^{-\beta (\epsk{k} - \muup_{N - 1})}
     \left( 1 - g_{k, N - 1} \right) \quad ; \quad
   \e^{\beta \muup_N}
   = \frac{N + 1}{\sum_k \e^{-\beta \epsk{k}}
     \left( 1 - g_{k, N} \right)}. 
   \label{eq:2Delgas:g_recur}
\end{equation}
While being valid for fermions with an arbitrary single-particle spectrum,
the recurrence relation (\ref{eq:2Delgas:g_recur}) turns out to
coincide~\footnote{Sch\"onhammer adopts the definition
$\muup_N = F_N - F_{N - 1}$, whereas we use $\muup_N = F_{N + 1} - F_N$.} 
with the one obtained by Sch\"onhammer in Eq.~(19) of
Ref.~\cite{Schoenhammer2000}, when applied to fermions with a linear energy
spectrum. Focusing on the linear energy spectrum, we note that the density
of the states in 2D wave vector space equals $L_x L_y / (2 \pi)^2$.
On average, a circle with radius $K$ thus encloses 
$n_K = \pi K^2 L_x L_y / (2 \pi)^2$ states, the single-particle energy on
the edge of the circle thus being
$\epsk{K} = \hbar^2 K^2 / (2m_e) = 2 \pi \hbar^2 n_K / (m_e L_x L_y)$.
For sufficiently large wave vectors, the single-particle energies can
therefore be replaced by a linearized spectrum
\begin{equation}
   \tilde{\epsilon}_n = n \Delta \quad ; \quad
   \Delta = \frac{2 \pi \hbar^2}{m_e L_x L_y} \quad ;
   \quad n = 0, 1, 2, \cdots, \label{eq:linspec}
\end{equation}
resulting in the following recurrence relation,
replacing~(\ref{eq:2Delgas:g_recur})
\begin{align}
   g_{n, N}^{\text{lin}}  
   & = \e^{-\beta \left( n \Delta - \muup_{N - 1}^{\text{lin}} \right)}
       \left( 1 - g_{n, N - 1}^{\text{lin}}\right),
       \label{eq:2Delgas:glin_recur} \\
   \e^{\beta \muup_N^{\text{lin}}}
   & = \frac{N + 1}{\sum_{n = 0}^{\infty}
       \e^{-\beta n \Delta} \left( 1 - g_{n, N}^{\text{lin}} \right)}.
       \label{eq:2Delgas:mulin_recur}
\end{align}
Clearly, the mere introduction of the linearized spectrum does not offer
any improvement on the numerical accuracy. The latter goal may be reached
most easily by implementing the analytical results obtained by
Sch\"onhammer~\cite{Schoenhammer2000}.
Although it is tempting to translate his formulas literally, some care is
required because he considers a spectrum
$\epsilon_i = i \Delta,~i = 1, 2, \cdots$. Of course, a gauge
transformation relates both approaches, but applying it in detail to all
intermediate relations and quantities is not a trivial task. Instead, a
careful recalculation adopting the notation of (\ref{eq:linspec}) and
following the approach we proposed in Sec.~4 of~\cite{WMLLFB2017}, yields
the following results for the CE partition function $Z_N^{\text{lin}}$, the
free energy $F_N^{\text{lin}}$, the internal energy $U_N^{\text{lin}}$ and
the chemical potential $\muup_N^{\text{lin}}$,
\begin{align}
   Z_N^{\text{lin}}
   & = \e^{-\beta N (N - 1) \Delta / 2} \prod_{n = 1}^N
       \frac{1}{1 - \e^{-\beta n \Delta}},
       \label{eq:2Delgas:Zlin} \\
   F_N^{\text{lin}}
   & = \frac{1}{2} N (N - 1) \Delta + \frac{1}{\beta}
       \sum \limits_{n = 1}^N \ln \left( 1 - \e^{-n \beta \Delta} \right),
       \label{eq:2Delgas:Flin} \\
   U_N^{\text{lin}}
   & = \left(
             \frac{1}{2} N (N - 1) + \sum \limits_{n = 1}^N
             \frac{n}{\e^{n \beta \Delta} - 1}
       \right)
       \Delta,
       \label{eq:2Delgas:Ulin} \\
   \muup_N^{\text{lin}}
   & = N \Delta + \frac{1}{\beta}
       \ln \left( 1 - \e^{-\beta(N + 1) \Delta} \right).
       \label{eq:2Delgas:mulin}
\end{align}
The analytical expression for $\muup_N^{\text{lin}}$ given by
Eq.~(\ref{eq:2Delgas:mulin}) not only replaces the numerical iteration
outlined in (\ref{eq:2Delgas:glin_recur}) and
(\ref{eq:2Delgas:mulin_recur}), but also enables the conversion
of~(\ref{eq:2Delgas:glin_recur}) into a recurrence relation connecting
subsequent level numbers $n$ for any \textbf{fixed} particle number~$N$:
\begin{align}
   g_{0, N}^{\text{lin}} = \,
   & 1 - \e^{-\beta N \Delta}, \label{eq:2Delgas:glin_0,N} \\
   g_{n + 1, N}^{\text{lin}} = \,
   & 1 - \e^{-\beta N \Delta} - 
     \frac{\e^{\beta \Delta (n + 1)} - 1}{\e^{\beta N \Delta}}
     g_{n, N}^{\text{lin}}, \label{eq:2Delgas:glin_n+1fromn,N} \\
   & \leftrightarrow~g_{n, N}^{\text{lin}} =
     \frac{\e^{\beta N \Delta}}{\e^{\beta \Delta (n + 1)} - 1}
     \left(1 - \e^{-\beta N \Delta} - g_{n + 1, N}^{\text{lin}}\right). 
     \label{eq:2Delgas:glin_nfromn+1,N}
\end{align}
as was already established by Sch\"onhammer~\cite{Schoenhammer2000}.
In order to remain fully self-contained, we (re)derive these equations in
the spirit of the projection operator approach in Appendix~\ref{appA}.
Subtle differences in the intermediate results as compared to, for
instance, Eqs.~(15,~20,~21) in~\cite{Schoenhammer2000} are due to the
energy scale (ground state energy $\epsk{0} = 0$ in the present
approach, but $\epsk{0} = \Delta$ in~\cite{Schoenhammer2000}), and the
above mentioned difference in the definition of the chemical potential.

The recurrence relation~(\ref{eq:2Delgas:glin_n+1fromn,N}) is numerically
accurate and stable as long as 
$\e^{\beta N \Delta} > \e^{\beta \Delta (n + 1)} - 1$, i.e.,
$n < \ln \left(\e^{\beta N \Delta} + 1 \right) / (\beta \Delta) - 1$.
However, even if $n$ is too large to fulfill this condition,
numerical convergence based on Eq.~(\ref{eq:2Delgas:glin_nfromn+1,N}) can
still be achieved, provided one finds a valuable \textit{initial} value
of $g_{n, N}^{\text{lin}}$, compatible with sufficiently large $n$.
The latter shouldn't be too difficult, since
$\lim_{n \to \infty} g_{n, N}^{\text{lin}} = 0$. Suppose that 
$g_{n_c + 1, N}^{\text{lin}}$ in~(\ref{eq:2Delgas:glin_nfromn+1,N}) is
negligible for some large enough $n_c$. Then $g_{n_c, N}^{\text{lin}}$
should satisfy $g_{n_{c}, N}^{\text{lin}} \lll 1$, i.e.,
$\e^{\beta \Delta \left( N - n_c -1 \right)} \lll 1$
which makes it capable of initializing~(\ref{eq:2Delgas:glin_nfromn+1,N}).
In practice, we required this condition to be satisfied in double precision
Fortran up to machine precision, but a less severe treatment should not
harm, since the numerical error in~(\ref{eq:2Delgas:glin_nfromn+1,N}) is
self-correcting. The main purpose of the present section being the
corroboration of our results by those obtained by Sch\"onhammer for the
linearized energy spectrum, we refer to~\cite{Schoenhammer2000} for a more
detailed investigation of the latter.

\section{Two-dimensional electron gas\label{sec:2DelgasQ} -- quadratic
         energy spectrum}
\begin{figure}[ht]
   \centering
   \includegraphics[width = 0.85\textwidth]{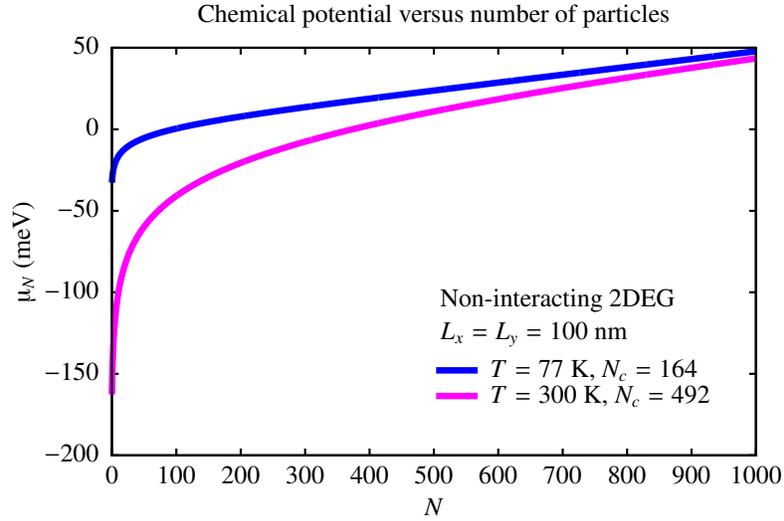}
   \caption{Chemical potential $\muup_N$ of a 2DEG at $T = 77, 300$ K as a
            function of the number of the particles $N$. $\Nc$ denotes the
            critical particle number that marks the cross-over between CE
            and GCE.}
   \label{fig:mu_TFX_N0-1000}
\end{figure}

\begin{figure}[ht]
   \centering
   \includegraphics[width = 0.85\textwidth]{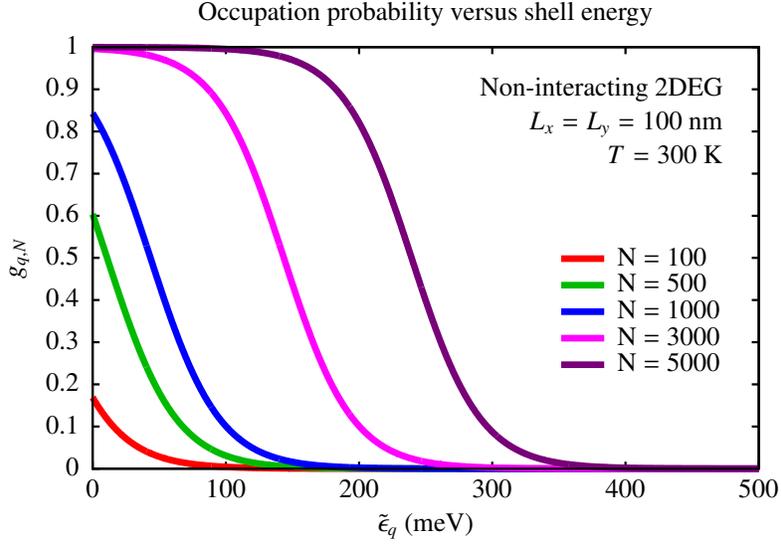}
   \caption{Electron occupation number (distribution function) versus the
            single-electron energy shell index $q$ calculated for 
            $T = 300$ K and for five values of $N$.}
   \label{fig:g_T300_NFX_q0-1032}
\end{figure}

Finally, we revisit the ordinary 2DEG, characterized by the quadratic
dispersion relation (\ref{eq:2DEGqspectrum}), and apply the numerical
algorithm, iterating on the recursion
relations~(\ref{eq:Fermi:g_recur} -- \ref{eq:Fermi:mu_recur}) for
$N \leqslant \Nc$ and avoiding the numerical instability issues for
$N > \Nc$, as outlined in
Eqs.~(\ref{eq:Fermi:f_O(1)} -- \ref{eq:pseudoGCE}).
As an illustration, we have shown the chemical potential as a function of
$N$ in Fig.~(\ref{fig:mu_TFX_N0-1000}) for $T = 77$ K and $T = 300$ K.
The figure also indicates the critical particle number $\Nc$ beyond which
the Taylor expansion based algorithm starts correcting the fermionic
occupation numbers that are found to violate the inequality
$0 \leqslant \tilde{g}_{q, N} \leqslant 1$. The precise value of $\Nc$ not
only depends on $T$ and the parameters that specify the single-electron
dispersion relation ($L_x$ and $L_y$ in the present case) but also on the
tolerance used to estimate the numerical errors on the occupation numbers.
The values of $\Nc$ reported in Fig.~(\ref{fig:mu_TFX_N0-1000}) correspond
to a tolerance of 10$^{-10}$. In addition, Fig.~(\ref{fig:mu_TFX_N0-1000})
clearly illustrates that the asymptotic, linear dependence on $N$ is
attained sooner at relatively low temperatures. The latter may be expected
from the closed-form expression -- Eq.~(32) in~\cite{WMLLFB2017} -- that is
available for the GCE chemical potential in the thermodynamic limit,
i.e. when $L_x, L_y, N \to \infty$ while the areal electron
concentration $\ns = N / (L_x L_y)$ remains finite:
\begin{equation}
   \muGCE_{\mathsf{TL}}
   = \frac{1}{\beta} \ln
     \left(
           \exp \left( \frac{2 \pi \beta \hbar^2 \ns}{m_e} - 1 \right)
     \right).
\end{equation}
Finally, the occupation numbers are plotted versus the shell energy
$\tilde{\epsilon}_q$ in Fig.~(\ref{fig:g_T300_NFX_q0-1032}) for different
values of $N$. Using the shell energy $\tilde{\epsilon}_q$ as the
independent variable instead of the very shell index $q$, we may
straightforwardly analyze the profile of the CE occupation numbers in
comparison with the Fermi-Dirac distribution that would govern a GCE
approach. It turns out that, for the adopted parameter set, the CE
distribution function profile is predominantly exponential up to $N = 500$,
while the deviation from a Fermi-Dirac distribution becomes negligible
for $N > 3000$.

\section{Conclusive remarks}
Not only the partition function and its derived quantities, but also the
boson and fermion occupation numbers (distribution functions) can be
extracted from a workable set of coupled recurrence relations that are
straightforwardly derived in the framework of the projection operator
approach. Except for the special case of one-dimensional harmonic
oscillators, analytical solutions of the recurrence relations are rare, if
not unavailable, and a numerical treatment turns out to be paramount for
most applications, especially in condensed matter physics and related
areas. \\
For bosons, one may accidentally have to deal with some minor
over/underflow related issues, but the numerical stability of the iterative
solutions is generally guaranteed thanks to the self-correcting nature of
the recurrence relations. \\
For fermions, the numerical errors on the occupation numbers are found
to grow rapidly beyond a critical value of the number of particles, as a
direct consequence of the well-known sign problem. However, the proximity
of the grand-canonical distribution function in that case was exploited
to construct a simple algorithm remedying the unstable steps in the
regime of large particle numbers. Moreover, a clear criterion assessing the
validity of this alternative algorithm has been established for practical
purposes.

\appendix{}
\section{Chemical potential of bosons -- upper limit \label{app:muNvseps0}}
This section demonstrates that, for any number of bosons, the chemical
potential cannot exceed the single-particle ground-state energy $\epsk{0}$,
i.e. $\muup_N(\beta) < \epsk{0}$ for all $N$. Equivalently, using the
identity
\begin{equation}
   \e^{\, \beta \, \muup_N(\beta)} = \frac{\ZN(\beta)}{Z_{N + 1}(\beta)},
\end{equation}
we must prove that
\begin{equation}
   Z_{N + 1}(\beta) > \e^{-\beta \epsk{0}} \ZN(\beta).
\end{equation}
To this end, we first introduce some auxiliary quantities:
\begin{align}
          u & = \e^{-\beta \epsilon_0}, \nonumber \\
        x_j & = Z_1(j \beta) - \e^{-j \beta \epsilon_0}
              = Z_1(j \beta) - u^j, \nonumber \\
   \Delta_j & = Z_j(\beta) - \e^{-\beta \epsilon_0} Z_{j - 1}(\beta)
              = Z_j(\beta) - u Z_{j - 1}(\beta), \qquad j = 1, 2, 3, \ldots
                \nonumber \\
   \Delta_0 & = 1.
   \label{eq:aux}
\end{align}
Note that, due to $Z_1(j \beta) > \e^{-j \beta \epsilon_0}$ for all
positive integer values of $j$, each $x_j$ is a strictly positive number.
With the above notation, it remains to be demonstrated that $\Delta_N > 0$ 
for $N \geqslant 1$. \\
First, we invoke mathematical induction to prove the identity
\begin{equation}
   \ZN(\beta) = \sum_{l = 0}^N u^l \Delta_{N - l}, \qquad
                N = 0, 1, 2, \ldots
   \label{eq:ZNDN}
\end{equation}
The latter trivially holds for $N = 0$ and $N = 1$ as can be seen by direct
application of (\ref{eq:aux}). Indeed, assuming that (\ref{eq:ZNDN}) holds
for all particle numbers up to $N > 1$, its validity for $N + 1$ boson
directly follows from
\begin{align}
   Z_{N + 1}(\beta)
   & = \Delta_{N + 1} + u \ZN(\beta)
     = \Delta_{N + 1} + u \left( \sum_{l = 0}^N u^l \Delta_{N - l} \right)
     \nonumber \\
   & = \Delta_{N + 1} + \sum_{l = 0}^N u^{l + 1} \Delta_{N - l}
     = \Delta_{N + 1} + \sum_{j = 1}^{N + 1} u^j \Delta_{N + 1 - j}
     = \sum_{l = 0}^{N + 1} u^l \Delta_{N + 1 - l}.
\end{align}
Next, we expand the defining expression of $\Delta_N$, using both
(\ref{eq:ZNDN}) and the recurrence relations for $\ZN(\beta)$ and
$Z_{N - 1}(\beta)$:
\begin{align}
   \Delta_N
   & = \ZN(\beta) - u Z_{N - 1}(\beta)
     = \frac{1}{N} \sum_{l = 1}^N Z_1(l \beta) Z_{N - l}(\beta)
      -\frac{u}{N - 1} \sum_{l = 1}^{N - 1} Z_1(l \beta)
       Z_{N - 1 - l}(\beta)
     \nonumber \\
   & = \frac{Z_1(N \beta)}{N} +
       \frac{1}{N} \sum_{l = 1}^{N - 1} Z_1(l \beta) Z_{N - l}(\beta)
      -\frac{u}{N - 1} \sum_{l = 1}^{N - 1} Z_1(l \beta)
       Z_{N - 1 - l}(\beta)
     \nonumber \\
   & = \frac{Z_1(N \beta)}{N} +
       \frac{1}{N (N - 1)} \sum_{l = 1}^{N - 1} Z_1(l \beta)
       \Bigl[ (N - 1) Z_{N - l}(\beta) - N u Z_{N - 1 - l}(\beta) \Bigr]
     \nonumber \\
   & = \frac{Z_1(N \beta)}{N} +
       \frac{1}{N (N - 1)} \sum_{l = 1}^{N - 1} Z_1(l \beta)
       \Bigl[
             (N - 1)
             \bigl( Z_{N - l}(\beta) - u Z_{N - 1 - l}(\beta) \Bigr)
            -u Z_{N - 1 - l}(\beta)  
       \Bigr]
       \nonumber \\
   & = \frac{Z_1(N \beta)}{N} +
       \frac{1}{N} \sum_{l = 1}^{N - 1} Z_1(l \beta)
       \Bigl[
             Z_{N - l}(\beta) - u Z_{N - 1 - l}(\beta)
            -u Z_{N - 1 - l}(\beta)
       \Bigr]
       \nonumber \\
   & = \frac{Z_1(N \beta) - u Z_{N - 1}(\beta)}{N} +
       \frac{1}{N} \sum_{l = 1}^{N - 1} Z_1(l \beta) \Delta_{N - l}
     = -\frac{u Z_{N - 1}(\beta)}{N} +
       \frac{1}{N} \sum_{l = 1}^N Z_1(l \beta) \Delta_{N - l}
       \nonumber \\
   & = -\frac{u Z_{N - 1}(\beta)}{N} +
       \frac{1}{N} \sum_{l = 1}^N x_l \, \Delta_{N - l} +
       \frac{1}{N} \sum_{l = 1}^N u^l \Delta_{N - l}
       \nonumber \\
   & = \frac{1}{N}
       \left[
             \sum_{l = 1}^N u^l \Delta_{N - l} - u Z_{N - 1}(\beta)
       \right]
       + \frac{1}{N} \sum_{l = 1}^N x_l \, \Delta_{N - l}
     = \frac{1}{N}
       \left[
             \ZN - \Delta_N - u Z_{N - 1}(\beta)
       \right]
       + \frac{1}{N} \sum_{l = 1}^N x_l \, \Delta_{N - l}
       \nonumber \\
   & = \frac{1}{N} \sum_{l = 1}^N x_l \, \Delta_{N - l}.
   \label{eq:ZNDNproof}
\end{align}
Clearly, since all $x_l$ and $\Delta_{N - l}$, appearing in the right-hand
side of (\ref{eq:ZNDNproof}), are strictly positive, we conclude that
$\Delta_N$ must be strictly positive as well, provided that $N \geqslant 1$.

\section{Fermion occupation numbers for a linear energy spectrum\label{appA}}
Inserting a linear energy spectrum
$\epsk{n}^{\text{lin}} = n \Delta,~n = 0, 1, 2, \cdots$ into
Eq.~(\ref{eq:Intro:Z_NinGtilde}), we obtain the generating function for
fermions $(\xi = -1)$ as
\begin{equation}
   \Gt(\beta, z) = \prod_{k = 0}^{\infty}
                   \left( 1 + z \e^{-\beta k \Delta} \right).
\end{equation}
From~(\ref{eq:Intro:Z_NinGtilde}) the corresponding partition function
becomes
\begin{equation}
   \ZN^{\text{lin}}
   = \frac{1}{2 \pi \im} \oint_{\left \vert z \right \vert > 0}
     \frac{1}{z^{N+1}}
     \prod_{k = 0}^{\infty} \left( 1 + z \e^{-\beta k \Delta} \right) \d z,
   \label{eq:appA:Z_integral}
\end{equation}
while the occupation number of level $n$ is derived
from~(\ref{eq:Occup:g_in_Gtilde}):
\begin{equation}
   g_{n, N}^{\text{lin}}
   = \frac{\e^{-n \beta \Delta}}{\ZN^{\text{lin}}}
     \frac{1}{2 \pi \im}
     \oint_{\left \vert z \right \vert > 0} \frac{1}{z^N}
     \prod_{k \geqslant 0, \neq n}^{\infty}
     \left( 1 + z \e^{-\beta k \Delta} \right) \d z,
   \label{eq:appA:g_integral}
\end{equation}
Consider first the ground state occupation $g_{0, N}^{\text{lin}}$. A
substitution $k = j + 1$ followed by a substitution
$z \e^{-\beta \Delta} = w$ gives
\begin{equation} 
   g_{0, N}^{\text{lin}}
   = \frac{\e^{-i(N - 1) \beta \Delta}}{\ZN^{\text{lin}}}
     \frac{1}{2 \pi \im}
     \oint_{\left \vert w \right \vert > 0} \frac{1}{w^{N}}
     \prod_{j = 0}^{\infty}
     \left( 1 + w \e^{-j \beta \Delta} \right) \d w.
   \label{eq:appA:gGS_integral}
\end{equation}
Replacing $j$ by $k$ and $w$ by $z$, and inspecting
Eq.~(\ref{eq:appA:Z_integral}), one immediately recognizes
$Z_{N-1}^{\text{lin}}$ in the right-hand side of
(\ref{eq:appA:gGS_integral}) such that 
$g_{0, N}^{\text{lin}} = \e^{-\beta (N - 1) \Delta}
Z_{N - 1}^{\text{lin}} / \ZN^{\text{lin}}$. Using~(\ref{eq:2Delgas:Zlin}),
one readily obtains 
\begin{equation}
   g_{0, N}^{\text{lin}} = 1 - \e^{-\beta N \Delta},
   \label{eq:appA:g_0,N}
\end{equation}
which is equivalent to the expression for
$\left \langle n_1 \right \rangle_N$ reported in Eq.~(15)
of~\cite{Schoenhammer2000}. Next, focusing on $n > 0$, we separate the
factor $1 + z$ corresponding to $k = 0$ from the infinite product in the
integral representation of $g_{n + 1, N}^{\text{lin}}$ to obtain:
\begin{equation}
   g_{n + 1, N}^{\text{lin}}
   = \frac{\e^{-n \beta \Delta}}{\ZN^{\text{lin}}}
     \frac{1}{2 \pi \im} \oint_{\left \vert z \right \vert > 0}
     \left( \frac{1}{z^N} + \frac{1}{z^{N - 1}} \right)
     \prod_{k = 1, \neq n + 1}^{\infty}
     \left( 1 + z \e^{-k \beta \Delta} \right) \d z.
\end{equation}
As for the case $n = 0$, we first make the substitutions
$k = j + 1, w = z \e^{-\beta \Delta}$, and rename them back again into
$w$ to $z$ respectively, to find
\begin{equation}
   g_{n + 1, N}^{\text{lin}}
   = \frac{\e^{-n \beta \Delta}}{\ZN^{\text{lin}}}
     \frac{1}{2 \pi \im} \oint_{\left \vert z \right \vert > 0}
     \left(
           \frac{\e^{-\beta (N - 1) \Delta}}{z^N} +
           \frac{\e^{-\beta (N - 2) \Delta}}{z^{N - 1}}
     \right)
     \prod_{k = 0, \neq n}^{\infty}
     \left( 1 + z \e^{-k \beta \Delta} \right) \d z.
\end{equation}
The contributions to the above integral corresponding respectively to the
fractions $\e^{-\beta (N - 1) \Delta} / z^N$ and
$\e^{-\beta (N - 1) \Delta} / z^{N - 1}$ are proportional to
$g_{n, N}^{\text{lin}}$ and $g_{n, N - 1}^{\text{lin}}$, as can be derived
from Eq.~(\ref{eq:appA:g_integral}). Hence, we obtain:
\begin{equation}
   g_{n + 1, N}^{\text{lin}}
   = \e^{-\beta N \Delta} g_{n, N}^{\text{lin}} +
     \e^{-\beta (N - 1)  \Delta} g_{n, N - 1}^{\text{lin}}
     \frac{Z_{N - 1}^{\text{lin}}}{\ZN^{\text{lin}}}.
\end{equation}
Using
$Z_{N - 1}^{\text{lin}} / \ZN^{\text{lin}} =
\left( 1 - \e^{-\beta N \Delta} \right) \e^{\beta (N - 1) \Delta}$ 
one rediscovers the recurrence relation Eq.~(18)
of~\cite{Schoenhammer2000}:
\begin{equation}
   g_{n + 1, N}^{\text{lin}}
   = \e^{-\beta N \Delta} g_{n, N}^{\text{lin}} +
     \left( 1 - \e^{-\beta N \Delta} \right) g_{n, N - 1}^{\text{lin}}~.
   \label{eq:appA:g_n,N}
\end{equation}
As such, the above recurrence relation is not particularly useful, with
both the energy level index and the particle number $N$ appearing as
incremental integers. However, the application of
(\ref{eq:2Delgas:glin_recur}) eliminating
$g_{n, N - 1}^{\text{lin}} = 1 - g_{n, N}^{\text{lin}}
\e^{\beta (n \Delta - \mu_{N - 1}^{\text{lin}})}$ and the use of
(\ref{eq:2Delgas:mulin}) finally yields a recurrence relation in $n$ only,
the value of $N$ remaining fixed,
\begin{equation}
   g_{n + 1, N}^{\text{lin}}
   = 1 - \e^{-\beta N \Delta} - \e^{-\beta N \Delta}
     \left( \e^{\beta \Delta (n + 1)} - 1 \right) g_{n, N}^{\text{lin}}~,
   \label{eq:appA:g_n_recur1}
\end{equation}
which is easily reversed from increasing to decreasing energy level index
$n$:
\begin{equation}
   g_{n, N}^{\text{lin}}
   = \frac{\e^{\beta N \Delta}}{\e^{\beta \Delta (n + 1)} - 1}
     \left( 1 - \e^{-\beta N \Delta} - g_{n + 1, N}^{\text{lin}} \right).
   \label{eq:appA:g_n_recur2}
\end{equation}

\section*{Acknowledgment}
The authors are indebted to Lucien Lemmens for useful discussions and
suggestions as well as for critical reading of the manuscript.

\section*{References}

\bibliography{CE-occnum-MB}

\end{document}